\documentstyle[12pt]{article}
\newcommand{\singlespace}{
    \renewcommand{\baselinestretch}{1}\large\normalsize}
\newcommand{\doublespace}{
    \renewcommand{\baselinestretch}{1.6}\large\normalsize}
\renewcommand{\thesection}{\arabic{section}}
\newcommand{\be}{\begin{equation}}
\newcommand{\ee}{\end{equation}}
\newcommand{\ba}{\begin{eqnarray}}
\newcommand{\ea}{\end{eqnarray}}
\newcommand{\ket}[1]{| {#1} \rangle}
\newcommand{\bra}[1]{\langle {#1} |}
\newcommand{\ave}[1]{\langle {#1} \rangle}
\newcommand{\mpi}{m_\pi}
\newcommand{\bpi}{\mbox{\boldmath $\pi$}}

\def\thefootnote{\fnsymbol{footnote}}
\newcommand{\footur}{\footnote[1]{\small also: University of Illinois at
Urbana-Champaign,
Urbana, IL 61801, USA \\}}
\newcommand{\footdu}{\footnote[2]{\small Permanent address: Joint Institute for
Nuclear Research,
Dubna, Russia}}

\begin{document}
\begin{titlepage}
\pagestyle{empty}
\vspace{1.0in}
\begin{flushright}
June 1994
\end{flushright}
\vspace{1.0cm}
\begin{center}
\doublespace

\begin{Large}
{\bf Temperature Dependence of the Chiral Condensate\\
from an Interacting Pion Gas}
\end{Large}
\vspace{1.0cm}

G. BUNATIAN{\footdu} and J. WAMBACH{\footur}\\
\vspace{0.2 in}
Institut f\"ur Kernphysik\\
Forschungszentrum J\"ulich\\
D-52425 J\"ulich, Fed. Rep. Germany

\end{center}
\vspace{1.5cm}

\begin{abstract}
By exploiting the fact that the chiral condensate is related to
the derivative of the free-energy density with respect to the
bare quark mass we calculate the temperature dependence of the
condensate ratio $\ave{\bar qq}_T/\ave{\bar qq}_0$ from
an iteracting pion gas. When using the Weinberg Lagrangian
at the Hartree level we find a depression of the condensate
with temperature. For $T < 100$ MeV the results are in
good agreement with chiral perturbation theory
in the three loop approximation. Near $T_c$, however,
there are marked differences due to non-perturbative nature
of our approach.
\end{abstract}
\end{titlepage}

\doublespace

\section{Introduction}

The aim of relativistic heavy-ion collisions with cm-energies $E_{cm}\ge
20$ GeV/A is to create a quark-gluon plasma at early stages of the collision.
It is predicted that the formation of the plasma is accompagnied by a chiral
restoration transition which, for two light flavors, is most likely second
order
\cite{Kars}. Much efford has been devoted to calculate this transition in
effective
models of QCD, in particular by using pions as effective degrees of freedom.
This approach exploits the fact that the chiral $SU_L(2)\times SU_R(2)$
symmetry is spontaneously broken in the non-perturbative vaccum
giving rise to nearly massless Goldstone bosons (pions) which dominate
the thermodymanics of the hadronic phase for temperatures $T\sim m_\pi$.
We adopt the simplified scenario of an interacting pion gas which is assumed to
be in thermal as well as chemical equilibrium, $\mu _{\pi} =0$, throughout the
time
evolution of the hadronic phase. This "black body radiation" will disappear
when
$T$
has dropped to zero and the pion field has reached its vacuum state.

To obtain an expression for the chiral condensate ratio
$\ave{\bar qq}_T/\ave{\bar qq}_0$ in terms of pionic degrees of freedom
we can use the fact that, according to the Feymann-Hellmann theorem,
$\ave{\bar qq}_T$ is related to the derivative of the free-energy density
(the thermodynamic potenial) $\tilde\Omega(T)$ with respect to the bare quark
mass $m$ as
\be
\ave{\bar qq}_T=\partial\tilde\Omega(T)/\partial m
\ee
(we can ignore the fact that up and down quark masses not the same).
Denoting the difference in free energy density as $\Omega(T)=\tilde\Omega(T)-
\tilde\Omega(0)$ and using the Gell-Mann Oakes Renner relation \cite{GeOR}
\be
\mpi^2f_\pi^2=-m\partial\tilde\Omega(0)/\partial m
\ee
one immediately derives that
\be
{\ave{\bar qq}_T\over\ave{\bar qq}_0}=
1-{1\over f_\pi^2}{\partial\Omega(T)\over\partial\mpi^2}
\ee
This expression is identical to that of ref.~\cite{GeLe} if one
ignores correction to $f_\pi$ and $\mpi$ of order $m$. Within the
current uncertainties their combined effect is only a few \% and
can be safely ignored in the present context.

In ref.~\cite{GeLe} chiral perturbation theory, including three loops, has been
used to evaluate $\Omega(T)$. As the authors point this
is equivalent to a virial expansion in powers of $T$ up to power $T^8$.
In the present paper we wish to persue a different route by calculating
$\Omega(T)$ from a non-pertubative method, well-known for phonon-phonon
interactions in condensed matter physics \cite{LuWa,CaPe}
 While less systematic
from the point of view of a gradient expansion of the effective QCD
Lagrangian it has the advantage of treating the vicinity of the phase
transition more properly.

\section{Free energy of a pion gas}

For the evaluation of the free energy $\Omega(T)$ we shall start from
well-known expression for interacting boson systems at finite temperature
\cite{LuWa,CaPe}. In terms of the pion Green's function \cite{AbGD}
\be
D(\xi,{\bf k} ,T)=(\xi^2-m_\pi^2-{\bf k} ^2-\Pi(\xi,{\bf k} ,T))^{-1},
\ee
the thermodynamic potential is given by
\ba
\Omega(T)=\Omega'(T)-{3\over 2}\int
{d{\bf k}\over (2\pi)^3}\int {d\xi\over \pi }
{\rm Im} \biggl \{
\rm{ln}[-D^{-1}(\xi,{\bf k} ,T)]
+D(\xi,{\bf k} ,T)\Pi(\xi,{\bf k} ,T)\biggr \}
\ea
where $\Omega'(T)$ is the sum of all contributions from
'skeleton diagrams' representing the perturbation
expansion of $\Omega$. These are evaluated by using full
single-particle Green's functions rather than bare ones.
Furthermore
\be
\delta \Omega /\delta \Pi =0
\ee
and
\be
\delta \Omega'/\delta D =\Pi
\nonumber\\
\ee
(see refs.\cite{LuWa,CaPe,AbGD}).
Since the physical pion mass enters the propagator $D$, the self energy
$\Pi (\xi ,{\bf k} ,T)$ is the difference
\be
\Pi(\xi,{\bf k},T)=\tilde\Pi(\xi,{\bf k},T)-\tilde\Pi(\xi,{\bf k},0),
\ee
whereby the infinite contributions to the mass operator
 $\tilde\Pi(\xi,{\bf k},0)$
due to vaccum fluctuations are removed from $\Pi (\xi ,{\bf k},T)$.
Consequently also $\Omega(T)$ is a {\it finite} difference
$\Omega(T)=\tilde\Omega(T)-\tilde\Omega(0)$ and no cut-off's are
needed.

The key quantity in eq.~(5) is the self energy $\Pi$. To calculate
it we use the Weinberg Lagrangian \cite {Wein}. To lowest order in the
coupling constant $1/f_{\pi}$ it takes the form
\ba
{\cal L}={\cal L}^0+{\cal L}'\qquad\qquad
\nonumber\\
{\cal L}^0={1\over 2}\partial_\mu\bpi\partial^\mu\bpi+{1\over 2}\bpi^2\mpi^2
\quad
\nonumber\\
{\cal L}'=\lambda(-\partial_\mu\bpi\partial^\mu\bpi+{1\over 2}\bpi^2\mpi^2)
\bpi^2
\ea
where $\lambda=f^{-2}_\pi/4$ ($f_\pi=93$ MeV).
Here ${\cal L}^0$ is free pion Lagrangian and ${\cal L}'$ denotes the
$\pi \pi -$ interaction. The second term in ${\cal L}'$, which
violates chiral symmetry for physical pions, is introduced
according to ref.~\cite{gur}. The Weinberg Lagrangian
reproduces the current algebra results for the $\pi \pi$ scattering lengthts,
and has been successful in describing low-energy $\pi \pi$ scattering
\cite{Lem}, which is sufficient in our case. The interaction Lagrangian
${\cal L}'$ is irreducible and describes a point-like pion-pion interaction.
Hence, by definition, it suffers no medium modification.

Given eq.~(8), the pion self energy $\tilde\Pi$ contains single-pion and
three-pion contributions
\be
\tilde\Pi(\xi,{\bf k} ,T)=
\tilde\Pi(\xi,{\bf k} ,T)^{1\pi}+\tilde\Pi(\xi,{\bf k} ,T)^{3\pi}
\ee
represented by the diagrams given in Fig.~1.
The lines represent the full propagator $D$. The small dots denote the
irreducible $\pi\pi$ vertex while the heavy dot indicates the full
$\pi\pi$ vertex \cite{AbGD}. Because of eq.~(8), the difference
$\Pi (\xi ,{\bf k},T)$ only includes intermediate virtual states with
energies $E\le T$. Because of resonances, the three-pion contribution
should reduce to diagrams involving heavy mesons in the intermediate states
(Fig.~2).  Since their masses
are larger than the temperature range of interest we shall only include
the first part of eq.~(10) which is equivalent to the
Hartree approximation.

In this case, the quantity $\Pi (\xi ,{\bf k},T)$ takes the form, see
ref.\cite{BuKa}
\be
\Pi ^{1\pi} (\omega ,{\bf k},T)=
-10d\lambda m_{\pi} ^2 +6d\lambda (\omega
^2 -{\bf k}^2)+6d\lambda \tilde \mpi ^2
\ee
and pion propagator is given by
\be
D(\omega ,{\bf k} ,T)
=\gamma (\omega ^2 -{\bf k}^2 -\tilde \mpi ^2 )^{-1},
\ee
It contains a temperature dependent effective mass
\be
\tilde\mpi^2={1-10\lambda d(T)\over 1-12\lambda d(T)}\mpi ^2
\ee
and a residue $\gamma=(1-6\lambda d(T))^{-1}$.
The quantity $d(T)$ is determined by \cite{BuKa}
\ba
d(T)={1\over 2\pi^2}\int_0^\infty {dk k^2\over \omega_k}
{\chi(\omega_k)\over 1-6\lambda d(T)}\quad;
\chi(\omega_k)=\biggl [\rm{exp}(\omega_k/T)-1\biggr ]^{-1}
\ea
with
\be
\omega_k^2={\bf k}^2+\tilde \mpi^2;\quad
\ee
It should be noted that, for any temperature, $d(T)$ never reaches
the value $1/12\lambda$.
In the Hartree approximation, $\Omega '$ is given by the diagram
displayed in Fig.~3 and takes the form
\ba
\Omega '={3\lambda\over 4}\int {d{\bf k} \over (2\pi )^3}
\int {d{\bf p} \over (2\pi )^3}
\int _{-\infty} ^{\infty} {d\xi\over \pi} \chi (\xi)
Im D(\xi ,{\bf k} ,T)\nonumber\\
\int_{-\infty}^{\infty}{d\eta \over \pi} \chi (\eta)
 Im D(\eta ,{\bf p} ,T)
(-10\mpi ^2 +6(\xi ^2 -{\bf k} ^2 )+6(\eta ^2 -{\bf p} ^2 ))\, .
\ea

The expressions in eq.~(5), eqs.~(11-15) and eq.~(16) determine
$\Omega(T)$ in the Hartree approximation and can be used to find
the chiral condensate ratio via eq.~(3). This is done in the next section.

\section{Chiral Condensate}

The thermodynamic potential $\Omega$ depends on the free pion mass
$\mpi$, both, explicitly and through the self energy $\Pi$. Due to the
stationarity condition (7) the  $\mpi$-dependence of $\Pi$ does
not influence the calculation of $\partial \Omega /\partial \mpi ^2$ and
we obtain
\ba
{\partial \Omega \over \partial \mpi ^2}={3\over 2}\int
{{d\bf k} \over (2\pi)
^3}\int_{-\infty} ^{\infty} {d\xi\over \pi}\chi (\xi )
 Im D(\xi ,{\bf k} ,T)
\nonumber\\-
{30\lambda\over 4}\int {{d\bf k} \over (2\pi)^3}\int
 {{d\bf p} \over (2\pi)^3}
\int_{-\infty}^{\infty} {d\xi\over \pi}\chi (\xi) Im D(\xi ,{\bf k},T)
\int_{-\infty}^{\infty} {d\eta \over \pi} \chi (\eta)
Im D(\eta ,{\bf p} ,T)
\ea
which can be expressed in terms of $d(T)$ as
\be
{\partial \Omega\over \partial \mpi ^2}={3\over 2}d(T)-{30\lambda
\over 4}d^2(T).
\ee
where $d(T)$ is the solution of the integral equation (14).
The result is displayed in Fig.~4. As expected, one observes a
decrease of $\ave{\bar qq}_T/\ave{\bar qq}$ with temperature
which, for the physical values of $\mpi$ MeV and $\lambda$,
reaches, however, a constant asymptotic value of $\sim 0.7$ at high
temperature. This is easily understood  from eq.~(18),
recalling the behaviour of the quantity $d(T)$ for
$T\to \infty$, as mentioned above. Our result is quite different from
the chiral perturbation theory \cite{GeLe}. In that
work the condensate ratio monotonically drops with temperature.
One should recall here that our calculations are carried out in the Hartree
approximation summing a selected, but infinite, number of loops. In such a
self-consistent scheme the strong pion-pion interaction first
enhances $\partial \Omega /\partial\mpi ^2$ with $T$ growth, but then it
prevents a further increase due to non-linearities.

To demonstrate the non-perturbative effects we have repeated the
calculation artificially increasing $f_\pi$ by a factor of two,
{\it i.e.} reducing the coupling constant by a factor of 4. In
this case perturbation theory should be more appropriate.
One finds that, for low values of $T$ the reduction, is the same
as with the full coupling, but then continues to higher $T$. Perhaps
accidentally these results are very close to those of ref.~\cite{GeLe}.

True chiral restoration only takes place in the chiral limit
$\mpi\to 0$. We have studied this limit as well and the result
is also shown in Fig.~4. In this case the temperature-dependent
effective mass $\tilde \mpi (T)$ remains zero and the pion propagator
becomes
\be
D_0 =\gamma (\omega ^2  -{\bf k} ^2)^{-1}.
\ee
with the residue $\gamma$ being of the same form as for $\mpi\neq 0$.
The solution for $d(T)$ in eq.~(14) now reduces to the very simple form
\ba
d_0 (T)={1-{\sqrt {1-24\lambda j(T)}}\over 12\lambda},\quad
\nonumber\\
j(T)=\int_0^{\infty}{k^2 dk\chi (\omega_k^0 )\over 2\pi ^2 \omega_k^0},
 \,\omega_k^0=k.
\ea
One concludes immediately that, for temperatures where $j(T)>1/24\lambda$
there no longer exists a real solution for $d_0(T)$, {\it i.e.} the residue
$\gamma$ becomes imaginary. Physically this means that, above the critical
temperature, $T_c$, where $j(T_c)=1/24$ no stable single-pion
states exist. This we interprete as the signature of the
chiral phase transition. At this point the chiral condensate remains
{\it finite}, however, due to non-pertubative effects. Reducing the
coupling strength to $\lambda/4$ (dashed line) the condensate drops
to zero again.

\section{Summary and Conclusions}

{}From the thermodynamic potential $\Omega(T)$ of an interacting pion
gas we have calculated the temperature dependence of the chiral condensate
in the Hartree approximation. As from other effective $QCD$ models,
we find a reduction with increasing temperature. In the chiral limit
we identify the critial temperature, $T_c$, for the chiral phase transition
as the point where the residue of the single-pion propagator becomes
purely imaginary. To within the numeircal accuracy   the well-known result,
$T_c=\sqrt{2}f_\pi$, the well-known result
from the linear $\sigma$ model. At this point, the chiral condensate
ratio is non-zero, however, with $\ave{\bar q q}_T\sim 0.7\ave{\bar q q}_0$
as a result of non-perturbative effects. By reducing the coupling constant
$\lambda$ in the chiral Lagrangian, we have shown, that perturbation theory
is valid at low temperatures $T < \mpi$ but breaks down near $T_c$. For the
physical pion mass $\ave{\bar qq}_T/\ave{\bar qq}_0$ drops more slowly
with $T$ and eventually reaches a constant value due to explicit
chiral symmetry breaking. This value is rather large.

For an improved description of the condensate ratio one should include
resonances such a the $\rho$-meson in the evalution of the pion self
energy. Furthermore a thorough treatment of the phase transition
demands proper account of the order-parameter fluctuations near
the critical point, as is well known. Work in this direction is
in progress.
\bigskip

\noindent
{\bf Acknowledgement}:
We thank G. Chanfray and M. Ericson for fruitful discussions.
G. Bunatian would like to thank the Deutsche Forschungsgemeinschaft
for support. This work is also supported in part by the National
Science Foundation under Grant No. NSF PHY89-21025.

\vfill\eject

\newpage
\begin{center}
{\Large \sl \bf Figure Captions}
\end{center}
\vspace{1.0cm}

\begin{itemize}
\item[{\bf Fig.~1}:] The pion self energy including 1-$\pi$ and 3-$\pi$
intermediate states.

\item[{\bf Fig.~2}:] Self-energy diagrams from heavy-meson exchanges.

\item[{\bf Fig.~3}:] Contribution to $\Omega'(T)$ in the Hartree approximation.

\item[{\bf Fig.~4}:]
Temperature dependence of the chiral condenstate from an interacting
pion gas. The full lines denote the results with $\lambda=f_\pi^{-2}/4$ both
for the physical pion mass and in the chiral limit. The dashed lines
give the corresponding results with $\lambda/4$.

\end{itemize}

\end{document}